\documentclass[twocolumn,aps,pra,superscriptaddress,showpacs]{revtex4-1}
\usepackage[latin9]{inputenc}
\usepackage{amsmath}
\usepackage{graphicx}
\usepackage{subfigure}

\makeatletter

\@ifundefined{textcolor}{}
{%
 \definecolor{BLACK}{gray}{0}
 \definecolor{WHITE}{gray}{1}
 \definecolor{RED}{rgb}{1,0,0}
 \definecolor{GREEN}{rgb}{0,1,0}
 \definecolor{BLUE}{rgb}{0,0,1}
 \definecolor{CYAN}{cmyk}{1,0,0,0}
 \definecolor{MAGENTA}{cmyk}{0,1,0,0}
 \definecolor{YELLOW}{cmyk}{0,0,1,0}

}
\newcommand{\ie}{{\it i.e.}}
\usepackage{bm}

\def\bra#1{\langle #1 \mid}
\def\ket#1{\mid #1\rangle}

\makeatother

\begin{document}

\title{The switching effect of the side chain on  quantum walks on triple graphs}

\author{Yi-Mu Du$^1$, Li-Hua Lu$^{1*}$ and You-Quan Li}

\affiliation{ Zhejiang Institute of Modern Physics and Department of Physics, \\
 Zhejiang University, Hangzhou 310027, P. R.  China}
 \email{lhlu@zju.edu.cn}
 \affiliation{
  Collaborative Innovation Center of Advanced Microstructures, Nanjing, P. R. China}

\begin{abstract}

We consider a  continuous-time quantum walk on a triple graph and  investigate  the influence  of the side chain on the propagation in the main chain. Calculating the interchange of the probabilities between the two parts of the main chain, we find that a switching effect appears if there are odd number of points on the side chain when concrete conditions between the length of the main chain and the position of the side chain are satisfied.   Whereas, such an effect does not occur if there are even number of points on the side chain.    We also suggest two proposals for experiment to observe such an effect, which may be employed to design new type of switching device.

\end{abstract}

\maketitle

\section{Introduction}
 Although the quantum walk was proposed  as the quantum mechanical counterpart of the classical random walk~\cite{Ahar}, it  exhibits exotic features  and has  extensive potential applications in many fields comparing with the classical random walk. For example, the quantum walk can not only provide us  a simple model to study the coherent quantum control over  atoms or photons in physical systems but also offer us  an advanced tool for new quantum algorithms~\cite{Moh, Shen, Sal}.  So the study of the quantum walk  has been receiving increasing attentions in recent years.  In the context of quantum information, the continuous-time and discrete-time  quantum walks were sequentially proposed~\cite{Ahar,Far} with the goal of applying them to quantum algorithms. For discrete-time quantum walks, one need a quantum coin to generate a superposition state for each step. Whereas, for continuous-time quantum walks, the quantum coin is not needed and the quantum-walk process is   realized  through the continuous tunneling between neighbor sites, which implies that the continuous-time quantum walk can be implemented on some possible lattice sites.

The propagation  features  of quantum walk is obviously affected by the graph where the quantum walk is implemented.  So far,  the quantum walks on different graphs,   such as hypercube~\cite{Kem,Kos,Kro,Mar,Adi}, cycle~\cite{c1,c2,c3}, hypercycle~\cite{hcy1} and percolation graphs~\cite{p1,p2,p3}, have been widely investigated.  For example, the mixing time and hitting time of quantum walks were studied in Ref.~\cite{Kem,Kos,Kro,Mar,Adi} where the hitting time of quantum walks for the opposite corners in  the hypercube graph  was shown to be exponentially faster than that  of classical random walks. The upper-bound of the mixing time of quantum walks on cycle graph was also estimated in Ref.\cite{c2}.  Additionally, in some literatures~ \cite{gi1,gi2,gi3,gi4}, the quantum walk was used to distinguish the isomorphism of the graph.

Since the quantum walks on different graphs can exhibit different characters, the study of that can not only  simulate some phenomena in conventional systems but also enlighten us to find more potential applications of quantum walk. Therefore we consider a  continuous-time quantum walk on a  triple  graph in this paper.  We show how  the length and the position of the side chain, and the length of  the main chain influence the propagation properties. Especially, we find that the quantum-walk system can exhibit obviously  different propagation properties when  the parity of the number of the points on the side chain is changed.  This parity effect on quantum walks is expected to be used for switching devices. A similar effect was noticed in mesoscopic metal ring~\cite{Ho} where there is a circulating current when a magnetic flux crosses the mesoscopic metal ring.  Such  a circulating current can be  influenced by the parity of the number of the electrons, which implies that the parity effect can  exhibit in mesoscopic metal microstructures. In a word, comparing  the quantum walk on one-dimensional lattices,  the existence of the side chain can obviously change the propagation properties of quantum walks, therefore it is worthwhile for us  to study the quantum walks on triple graphs.

 This paper is organized as follows. In the next Section, we present the model of the system.  In Sec.~\ref{subsec:onepoint},  with the help of Green function we investigate the quantum walk on the triple graph with the side chain of one point and find that the system can exhibit the switching effect  via changing the length of the main chain or the position of the side chain when the tunneling strength on the side chain is much larger than that on the main chain. In Sec.~\ref{subsec:twopoint} and Sec.~\ref{subsec:morepoint}, we investigate the case of the side chain with two points and more points, respectively. We find  that when the number of points on the side chain is even, the system can not exhibit the switching effect no matter what the length of the main chain is, which is contrast to the case of the side chain with odd points. In  the last section,  we propose two possible  experimental schemes to observe or apply the dynamical properties exhibited in the system we considered and give a brief summary.

\section{Modelling quantum random walk on triple graph}

We consider quantum random walks of single particle on the following triple type graph,
\[
\begin{picture}(66,79)(38,-59)
\setlength{\unitlength}{3.3mm}\linethickness{0.8pt}
\put(0,0){\circle*{0.6}} \put(0,0){\line(1,0){2}} \put(2,0){\circle*{0.6}}
\put(2,0){\line(1,0){0.8}}
\put(3.2,0){\line(1,0){0.2}} \put(3.6,0){\line(1,0){0.2}} \put(4,0){\line(1,0){0.2}}
\put(4.6,0){\line(1,0){0.8}}
\put(5.4,0){\circle*{0.6}} \put(5.4,0){\line(1,0){2}} \put(7.4,0){\circle*{0.6}} \put(7.4,0){\line(1,0){2}} \put(9.4,0){\circle*{0.6}}
\put(9.4,0){\line(1,0){0.8}}
\put(10.6,0){\line(1,0){0.2}} \put(11,0){\line(1,0){0.2}} \put(11.4,0){\line(1,0){0.2}}
\put(12,0){\line(1,0){0.8}}
\put(13.1,0){\circle*{0.6}}
\linethickness{0.6pt}
\put(7.4,-2){\circle*{0.6}} \put(7.36,-2){\line(0,1){2}}\put(7.36,-2){\line(0,-1){0.8}}
\put(7.36,-3.2){\line(0,-1){0.2}}
\put(7.36,-3.6){\line(0,-1){0.2}}
\put(7.36,-4.0){\line(0,-1){0.2}}
\put(7.36,-4.5){\line(0,-1){0.8}}
\put(7.4,-5.4){\circle*{0.6}}
\end{picture}
\]
The graph contains a main chain of $N$ points and a side chain of $S$ points
where the side chain is connected to a certain point on the main chain.
It becomes more convenient for us to label the points on the main chain by $1$, $2$, $\cdots$, $N$,
and the points on the side chain by $N+1$, $N+2$, $\cdots$, $N+S$.
Then the graph can be equivalently plotted as
\[
\begin{picture}(66,63)(68,-28)
\linethickness{0.8pt}
\setlength{\unitlength}{3.3mm}
\put(0,0){\circle*{0.6}}\put(-0.3,1){$\tiny{1}$}
\put(0,0){\line(1,0){2}}
\put(2,0){\circle*{0.6}}\put(1.7,1){$2$}
\put(2,0){\line(1,0){0.8}}
\put(3.2,0){\line(1,0){0.2}} \put(3.6,0){\line(1,0){0.2}} \put(4,0){\line(1,0){0.2}}
\put(4.6,0){\line(1,0){0.8}}
\put(5.4,0){\circle*{0.6}} \put(5.4,0){\line(1,0){2}}
\put(7.4,0){\circle*{0.6}} \put(7.1,1){$\ell$} \put(7.4,0){\line(1,0){2}}
\put(9.4,0){\circle*{0.6}}
\put(9.4,0){\line(1,0){0.8}}
\put(10.6,0){\line(1,0){0.2}} \put(11,0){\line(1,0){0.2}} \put(11.4,0){\line(1,0){0.2}}
\put(12,0){\line(1,0){0.8}}
\put(13.1,0){\circle*{0.6}}\put(12.7,1){$N$}
\linethickness{0.6pt}
\put(15,0){\circle*{0.6}} \put(14.4,1){$N\!+\!1$}
\put(15,0){\line(1,0){2}} \put(17,0){\circle*{0.6}}
\put(17,0){\line(1,0){0.8}}
\put(18,0){\line(1,0){0.2}} \put(18.4,0){\line(1,0){0.2}} \put(18.8,0){\line(1,0){0.2}}
\put(19.2,0){\line(1,0){0.8}}
\put(20.3,0){\circle*{0.6}}\put(19.7,1){$N\!+\!S$}
\put(11.1,0){\oval(7.6,3.3)[b]}
\end{picture}
\]
That is a one-dimensional chain of $N+S$ points with a connection broken between points $N$ and $N+1$
while with an additional $^\prime$long-distance$^\prime$ connection between points $\ell$ and $N+1$.
This implies that the  point of $\ell$ is a special point that divides the main chain into two parts.
Then we can model the quantum random walk on the triple type graph by the following Hamiltonian
\begin{eqnarray}
H &=& \Bigl(- \sum_{j=1}^{N-1}\ket{j}\bra{j+1} - J\sum_{j=N+1}^{N+S-1}\ket{j}\bra{j+1}
      \nonumber \\[2mm]
  & & -J\ket{\ell}\bra{N+1}\Bigr)  + ~\mathrm{h.c.},\label{eq:model}
\end{eqnarray}
where $\ket{j}$ denotes the state that a particle occupies the $j$-th site on the one-dimensional lattice
and $J$  the hopping strength on the side chain and that between the side chain and the $\ell$-th point of the main chain. The hopping strength on the main chain is set to unit for simplicity.

We know that the properties of continuous-time quantum walk sensitively depend on the type of graphs where the quantum walk is implemented. Although the quantum walk in one dimension  has well studied, the add of the side chain can obviously change the propagation character of quantum walks. Then our main purpose is to find out how the side chain influences the
state evolution on main chain by calculating of the interchange of the
probabilities between the two parts of the main chain. And we further expect to use the side chain to manipulate the evolution  of the probability  on the main chain, which maybe enlightening for switching devices.

\section{Green Function and Dynamical propagation features}

Let us start from the main-chain relevant part of the total Hamiltonian (\ref{eq:model}),
\[
H_\textrm{M} = -\sum_{j=1}^{N-1}\ket{j}\bra{j+1} + ~\mathrm{ h.c.}
\]
We know that its eigenvalues and the eigenstates  are given by
\begin{eqnarray}\label{eq:eigenvalues}
E_m &=& -2\cos\left( m\frac{\pi}{N+1} \right),
 \nonumber\\[2mm]
\ket{\psi_m} &=& \sum_{j=1}^N\sqrt{\frac{2}{N+1}}\sin\left(j m \frac{\pi}{N+1} \right)\ket{j}.
\end{eqnarray}
Here, $m=1$, $2$, $\cdots$, $N$, and $j=1$, $2$, $\cdots$, $N$.
Then we can obtain the Green function of the main chain Hamiltonian $H_{M}$
\begin{equation}\label{eq:go}
g_{0}(z)=\sum_{m=1}^{N}\frac{\ket{\psi_m}\bra{\psi_m
}}{z + 2\cos(\frac{m\pi}{N+1})}.
\end{equation}

Note that when  the side chain is included, the Green function of our model (\ref{eq:model}) can also  be obtained. Then with the help of Green function and numerical calculation,  we are mainly aim to find out   how the side chain influences the
state evolution on main chain by calculating of the interchange of the probabilities between the two parts of the main chain.  Here the points in the left  of the connection  point $\ell$ (\ie, the points $j$ with $1\leq j<\ell$) are consist of the left part,  and the other points in the main chain  (\ie, the points $j$ with $\ell<j\leq N$) are consist of the right part. Taking the two cases of $S=1$ and $S=2$ as examples (\ie, there are one or  two sites on the side chain, respectively ),  we find that the propagation features for $S=1$ and $S=2$ are completely different even if the other parameters and initial states are the same. Additionally, we  also show the influence of the position of the connection  point $\ell$ and the number of points in the main chain  on the propagation properties of the system.

\subsection{Side chain with one point}\label{subsec:onepoint}
Now we are in the position to consider the case of side chain with one point (\ie, $S=1$). In this case, the Green function can be written as
\begin{eqnarray}\label{eq:grf1}
G_{1} &=& \frac{\ket{N+1}\bra{N+1}}{z}+g_{0}+\frac{g_{0}\ket{\ell}\bra{\ell}g_{0}}{z/J^{2}-\left<\ell|g_{0}|\ell\right>}
\nonumber \\[2mm]
 && +\frac{\bra{\ell}g_{0}\ket{\ell}}{z}\times\frac{\ket{N+1}\bra{N+1}}{z/J^{2}-\left<\ell|g_{0}|\ell
 \right>}
 \nonumber \\[2mm]
 && -\frac{1}{J}\times\frac{g_{0}\ket{\ell}\bra{N+1}+\ket{N+1}\bra{\ell}g_{0}}{z/J^{2}-\left<\ell|g_{0}
 |\ell\right>}.
\end{eqnarray}
From the above Green function, one can easily obtain the eigenvalues of the  Hamiltonian~(\ref{eq:model}) for $S=1$. We  find that the existence of the side chain with one point divides  the eigenvalues of the main chain Hamiltonian $H_\mathrm{M}$ into two kinds. That  eigenvalues  of $H_{\mathrm{M}}$ remain when  the corresponding eigenstates satisfy that $\langle \ell|\psi_m\rangle=0$ (\ie, such eigenvalues are also eigenvalues of  the  Hamiltonian~(\ref{eq:model}) for $S=1$),  while  the others are shifted and  no longer the eigenvalues of the Hamiltonian~(\ref{eq:model}) for $S=1$ due to the existence of the side chain. Except for the remaining eigenvalues of $H_\mathrm{M}$, the other eigenvalues of the  Hamiltonian~(\ref{eq:model}) for $S=1$ can be obtained  via the roots of equation $z/J^{2}-\left<\ell|g_{0}|\ell\right>=0$.

Let us see the case of $J\gg 1$ where the above  equation can be solved analytically, and the corresponding roots read
\begin{eqnarray}\label{eq:root}
&&z=\begin{cases}
 & z_{0}+\triangle(z_{0})\\[1mm]
 & \pm J
\end{cases},\nonumber\\
&&\triangle(z_{0})=-\frac{z_{0}}{J^{2}\left<\ell|g_{0}(z_0)^{2}|\ell\right>},
\end{eqnarray}
where $z_0$ is determined by $\left<\ell|g_{0}(z_{0})|\ell\right>=0$.
Substituting Eq.~(\ref{eq:go}) into Eq.~(\ref{eq:root}), we can obtain the value of $z_0$. Then the changed energy levels can be written as
\begin{eqnarray}
E_n&=&-2\cos(\frac{n\pi}{\ell}),\\
E_{n'}&=&-2\cos(\frac{n'\pi}{N+1-\ell}),
\end{eqnarray}
with $n=1, 2, \cdots \ell-1$ and $ n'=1, 2, \cdots N-\ell$.  Here
the perturbation term $\Delta(z_0)\sim\frac{1}{J^2}$ are neglected due to the fact $J\gg 1$. From the above expressions of $E_{n(n')}$, we can find that if $(N+1)$ and $\ell$ have the greatest common divisor larger than two,  some energy levels of $E_n$ can be degenerate with that of  $E_{n'}$, respectively,  and the corresponding values of that degeneracy  energy levels  are just  equal to those remaining eigenvalues of $H_\mathrm{M}$. Whereas, if $(N+1)$ and $\ell$ have no the greatest common divisor larger than two, once the side chain with one point is included, none of the eigenvalues of $H_\mathrm{M}$ remain and the energy levels of $E_n$ and $E_{n'}$ are not degenerate. The above expression  implies that the relation between  the position of the connection  point and the number of points on the main chain can obviously affect the distribution  of the energy levels of the system. So the propagation properties in the triple graph we considered can exhibit essentially different features for different numbers of points on main chain and the  positions of the connection point.

\begin{figure}[ht]\vspace{3ex}
\includegraphics[height=45mm]{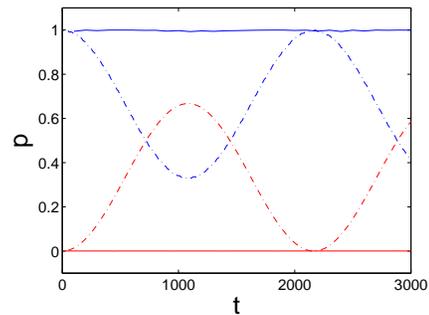}
\caption{(color online) The time evolution of the probabilities for the particle in the two different parts of the main chain, respectively,  for $S=1$. The parameters are $J=10$, $N=11$,  and $\ell=5$ for the solid lines and $\ell=6$ for the dot lines.   }
\label{fig:pros1}
\end{figure}

In order to investigate the propagation  features of the particle in the triple graph, let us observe  the time evolution of probability amplitude.
Assuming the particle is in the $j_2$-th site of the main chain, one can write  the probability amplitude $A(j_1,j_2)$  for the particle being in the $j_1$-th site at the time $t$ as
\begin{eqnarray}
 A(j_1,j_2)= \underset{E}{\sum}\mathrm{Res}G_{1}(E,j_{1},j_{2})\exp(-iEt),
\end{eqnarray}
where $G_{1}(E,j_{1},j_{2})=\bra{j_1}G_1(E)\ket{j_2}$ with  $E$ including all the eigenvalues  of the Hamiltonian~(\ref{eq:model}) for $S=1$, and the sign  $'\mathrm{Res}'$ stands for the residue of a function. If $j_1>\ell$ and $j_2<\ell$,  the probability amplitude can be given as
\begin{eqnarray}\label{eq:amplitude}
&&p(j_1,j_2) =O(\frac{1}{J^{2}})+\nonumber\\[2mm]
&&\underset{E\in E_{r}}{\sum}\frac{2}{N+1}\sin[j_{1}\theta(E)]\sin[j_{2}\theta(E)]e^{(-iEt)}[e^{(-i\triangle(E) t)}-1],\nonumber\\
\end{eqnarray}
where $E_r$ refer to  the remaining eigenvalues  of $H_{\mathrm{M}}$ ( \ie, $E_r$ are the energies given in Eq.~(\ref{eq:eigenvalues}) whose corresponding eigenstates  $\ket{\psi_m}$ satisfy  $\langle \ell|\psi_m\rangle=0$),  $\theta(E)=\displaystyle\arccos(-\frac{E}{2})$ and the expression of $\Delta(E)$ is given in Eq.~(\ref{eq:root}).  From the above expression~(\ref{eq:amplitude}), we can find that for $J\gg 1$, the value of $p(j_1,j_2)$ is not zero only when there are non-zero eigenvalues  of $H_{\mathrm{M}}$ remained, \ie,  $(N+1)$ and $\ell$ have the greatest common divisor larger than two.  So that the particle can cross the connection point from the one part into the other part of the main chain for $(N+1)$ and $\ell$ having the greatest common divisor larger than two  while the particle  is always in the initial part of the main chain for $(N+1)$ and $\ell$ having no  greatest common divisor larger than two, which will be  confirmed by Fig.~\ref{fig:pros1} . Additionally, we can also obtain  the probability amplitude of particle from the main chain into the side chain
\begin{equation}
\mathrm{Res}\langle j|G_{1}(J)|N+1\rangle=-\frac{1}{2}\delta_{j\ell}-\frac{1}{2J}\langle j|H_{\mathrm{M}}|\ell\rangle,
\end{equation}
which is zero as long as the particle is not in the connection point at the initial time.

In Fig.~\ref{fig:pros1}, we plot the time evolution of  probabilities  for particles in the left part and in the right part of the main chain, respectively, for different values of $\ell$. Note that in this and the following figures, the unit of time $t$ is in the unit of the inverse  of the hopping strength on the main chain, and that of tunneling strength $J$ is in the unit of the hopping strength on the main chain. At the initial time, the particle is in the third point   so that the particle is initially in the left part of the main chain.   Since the probability amplitude $A(j_1,j_2)$ for the particle tunneling between different points is given in Eq.~(\ref{eq:amplitude}),   the probability of the  particle in  the right part of the main chain at the time $t$ can be  written as $\sum_{\ell<j_1\leq N}|A(j_1,3)|^2$, and that in the left part of the main chain is $\sum_{1\leq j_1<\ell}|A(j_1,3)|^2$.    Note that once the particle is in the left part of the main chain, which concrete points that the particle is in does not affect the evolution  features of the probability distribution  in the two parts of the main chain. This is also  true for the case of side chain with more than one points.  From Fig.~\ref{fig:pros1}, we can find that the particle for $\ell=6$ ($N+1$ and $\ell$ having the greatest common divisor larger than two), the probabilities in the left and right parts of the main chain oscillate quasi-periodically with time.  Whereas,  the particle is always stay in the left part of the main chain and  has no probability to cross the connection point $\ell$ into the right part of the main chain for $\ell=5$ ($N+1$ and $\ell$ having no greatest common divisor larger than two), which exhibits the switching effect of the side chain. Note that, except for the position of the connection point (\ie, the value of $\ell$), the other parameters and the initial states are all the same for the two cases.


\subsection{Side chain with two points}\label{subsec:twopoint}
\begin{figure}[ht]
\vspace{3ex}
\subfigure{\includegraphics[width=55mm]{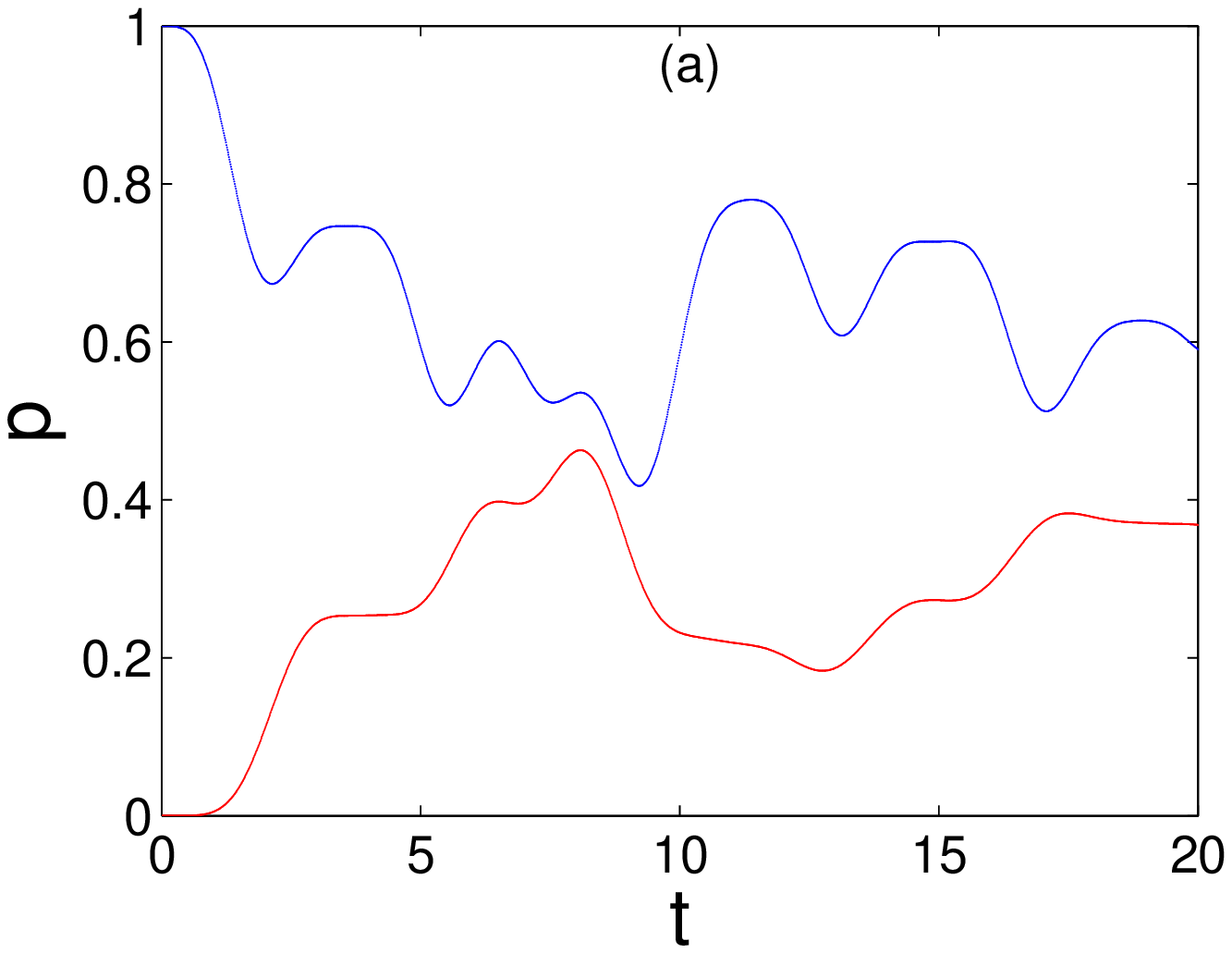}}
\subfigure{\includegraphics[width=55mm]{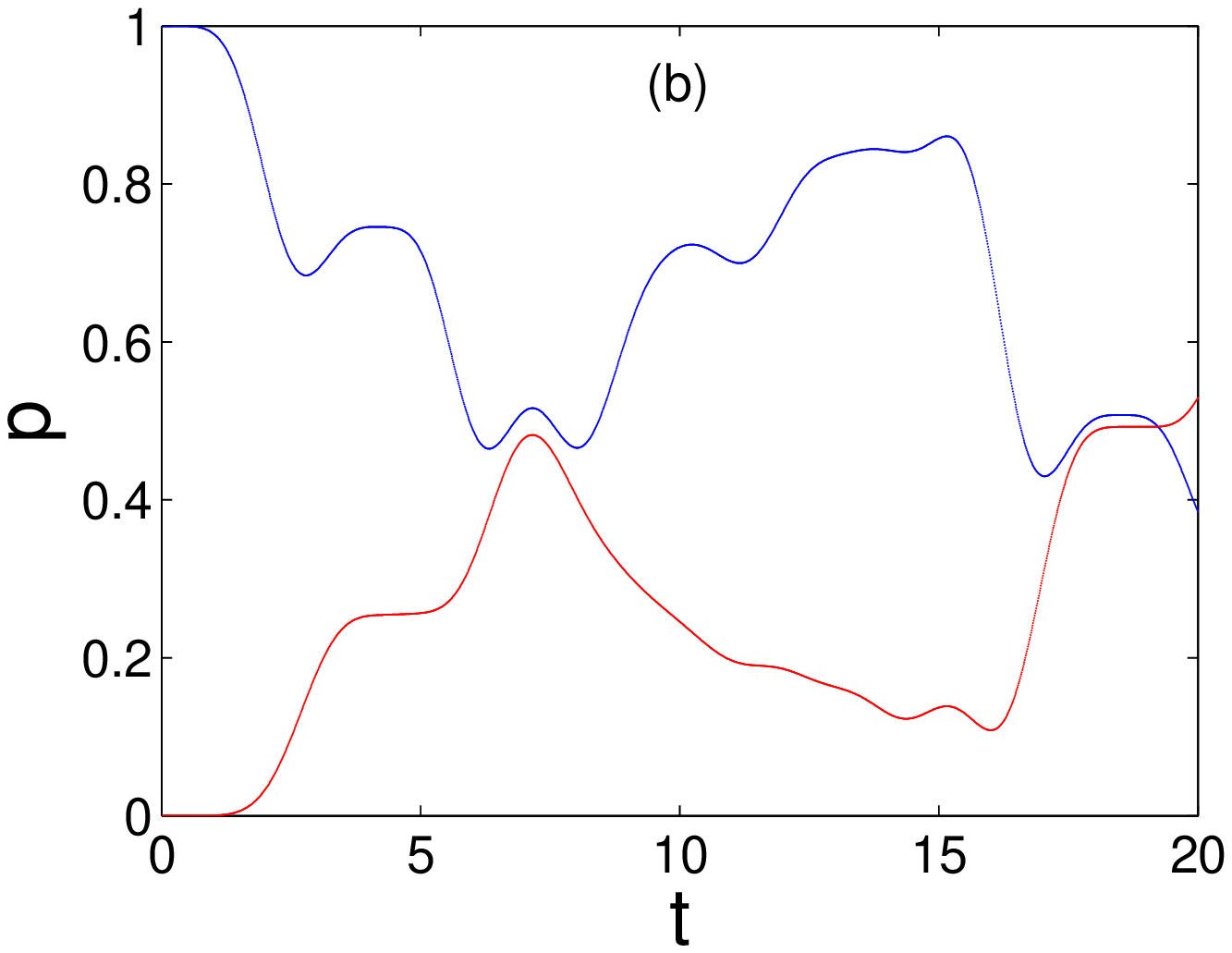}}
\protect\protect\protect\caption{(color online) The time evolution of the  probabilities for the particle in the two different parts of the main chain for $S=2$. The parameters are $J=10$, $N=11$,  and $\ell=5$ (a), $\ell=6$ (b). }
\label{fig:twopoint}
\end{figure}

For the Hamiltonian~(\ref{eq:model}) with $S=2$,  the Green function of the system is written as $G_2$ whose expression can be given carefully. Then the energy eigenvalues of the system  can also be obtained with the help of $G_2$. For the case of $J\gg 1$, some energy levels of $H_{\mathrm{M}}$  also remain if $(N+1)$ and $\ell$ have the greatest common divisor larger than two, and the others are determined via the roots of
\begin{equation}\label{eq:root2}
\frac{1}{z}+\bra{\ell}g_0\ket{\ell}=0,
\end{equation}
 and are not degenerate no matter whether $(N+1)$ and $\ell$ have the greatest common divisor larger than two.
In analogy to the discussion in the above subsection, we can obtain the probability amplitude of a particle from one point into another point on the main chain by calculating the residues of
\begin{equation}\label{eq:residue}
\bra{j_{1}}G_{2}\ket{j_{2}}=\bra{j_{1}}g_{0}\ket{j_{2}}+\frac{\bra{j_{1}}g_{0}\ket{\ell}\bra{\ell}g_{0}\ket{j_{2}}}{z/J^{2}
-\bigl(\displaystyle\frac{1}{z}+\left<\ell|g_{0}|\ell\right>\bigr)}.
\end{equation}
Although it is difficult to give the analytical expression of the residues of Eq.~(\ref{eq:residue}), we can find that the summation of such  residues is always not zero for any values of $(N+1)$ and $\ell$ because  the non-zero energy levels  near zero are also  existed for the system with the side chain of two points. That implies that the particle can always cross the connection point into the other part of the main chain no matter whether $(N+1)$ and $\ell$ have the greatest common divisor larger than two, which can be confirmed by Fig.~\ref{fig:twopoint}.

   In Fig.~\ref{fig:twopoint}, we plot the time evolution of the probability of the particle in the left and right part of the main chain, respectively.  Note that  we find that  the probabilities  of the particle in the left and right part of the main chain oscillate  so fast through the numerical calculation.  So in order to make Fig.~\ref{fig:twopoint} clear, we take a smaller time comparing to Fig.~\ref{fig:pros1}.  At the initial time, the particle is in the left part of the main chain. From this figure, we can find that the probabilities of the particle in the two parts of the main chain oscillate with time for both $\ell=5$ and $\ell=6$, which is in contrast to the case of side chain with one point (see Fig.~\ref{fig:pros1}). Additionally, the oscillation of the probability is not qusi-periodical like that in Fig.~\ref{fig:pros1} due to the fact that the roots of Eq.~(\ref{eq:root2}) are not periodical.

\subsection{Side chain with more points}\label{subsec:morepoint}

\begin{figure}[ht]
\vspace{3ex}
\subfigure{\includegraphics[width=55mm]{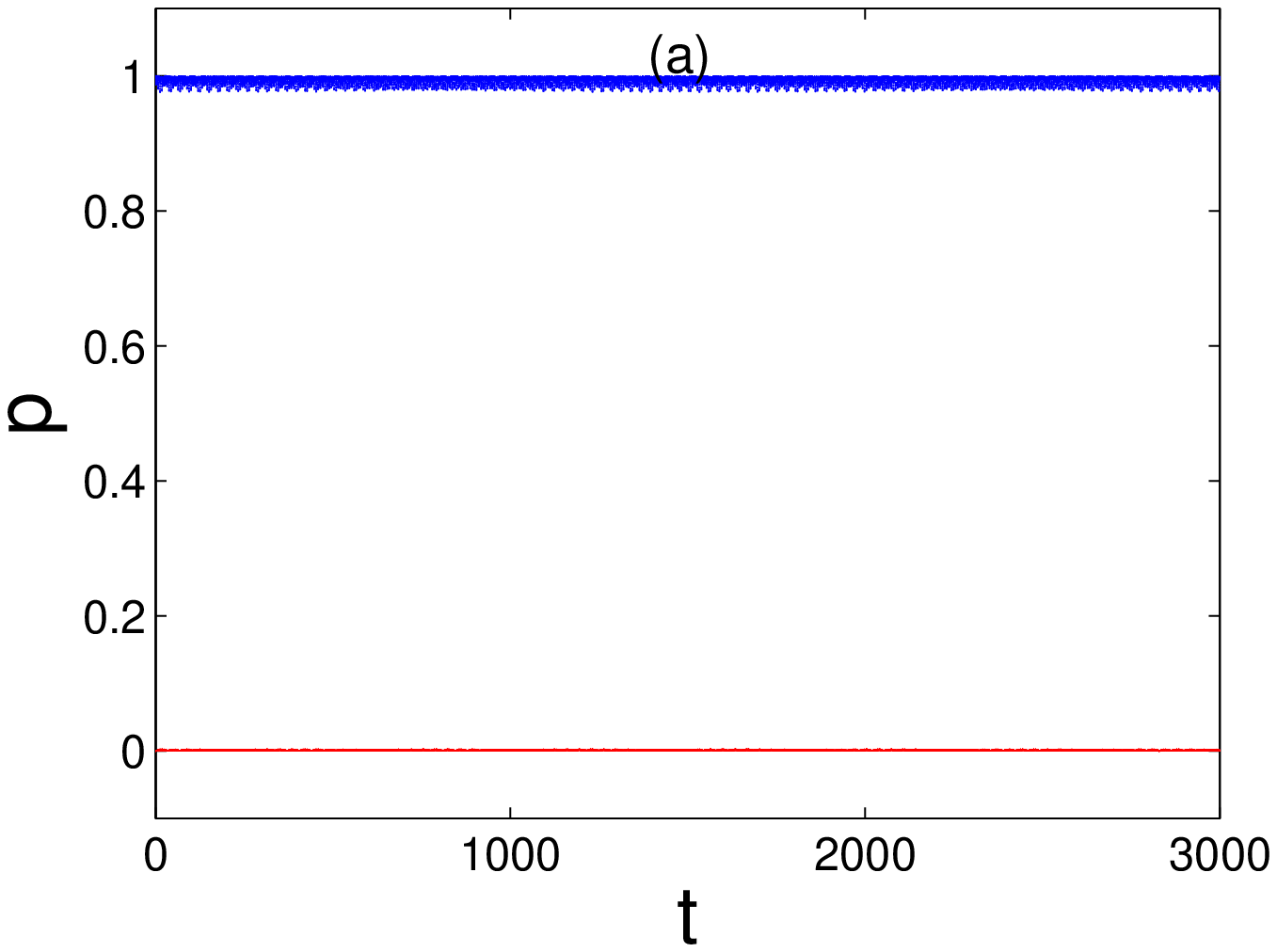}}
\subfigure{\includegraphics[width=55mm]{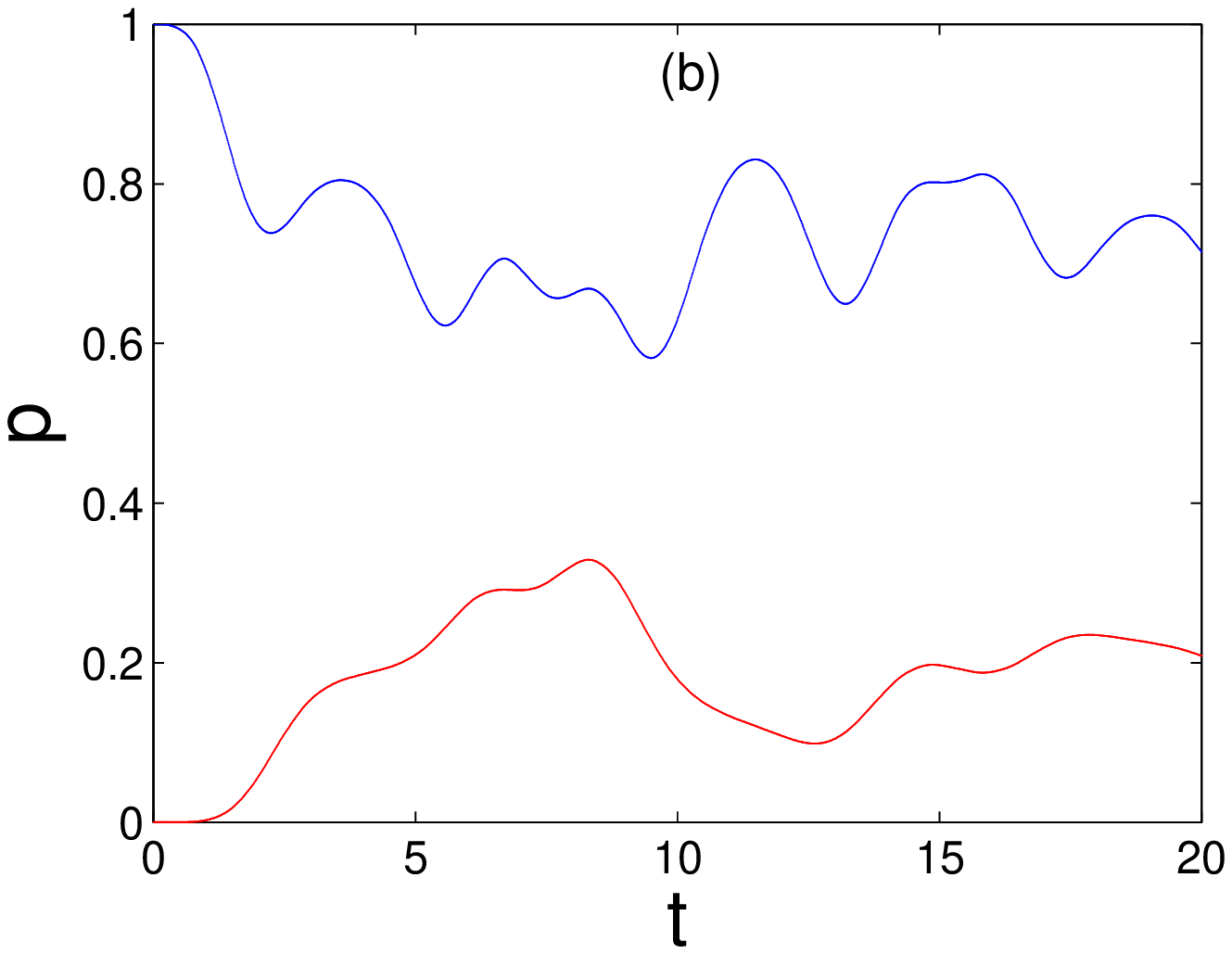}}
\protect\protect\protect\caption{(color online) The time evolution of the probabilities  for the particle in the two different parts of the main chain for different values of $S$. The parameters are $\ell=5$, $J=10$, $N=11$, and $S=3$ (a), $S=4$ (b).}
\label{fig:morepoint}
\end{figure}

In the above subsections, we consider two examples of side chain with one and two points, respectively. Now, let us see the case of the side chain with more than two points. In Fig.~\ref{fig:morepoint}, we plot the time evolution of the probability distribution in the left and right parts of the main chain, respectively, for different values of $S$.  Here the value of $\ell=5$, \ie., $N+1$ and $\ell$ have no greatest common divisor larger than two,  and the particle is in the left part of the main chain at the initial time. From this figure, we can find that the particle can not cross the connection point into the other part of the main chain for $S=3$, which is in contrast to  the cases of  $S=4$. Meanwhile, comparing the Fig.~\ref{fig:morepoint} (a) with Fig.~\ref{fig:pros1} , and Fig.~\ref{fig:morepoint} (b) with Fig.~\ref{fig:twopoint} (a) , respectively, we can see that the system of the side chain with odd points (even points) exhibits  analogous propagation features.  But the features of the system with  side chain of odd points differ significantly from that of the system  with side chain of  even points.   Such a feature  can be regarded as the parity effect in the quantum walk, which can be explained with the help of Green function. The parity effect also appeared in the other  physical systems, such as the mesoscopic metal ring~\cite{Ho}.

For the system of side chain with more points, the Green function of the system can be written as
\begin{eqnarray}
G_{S}&= & \overset{\sim}{g}_{0}+g_{0}+\frac{g_{0}\ket{\ell}J^{2}\times\bra{N+1}\overset{\sim}{g}_{0}\ket{N
+1}\bra{\ell}g_{0}}{1-J^{2}\times\bra{N+1}\overset{\sim}{g}_{0}\ket{N+1}\bra{\ell}g_{0}
\ket{\ell}}\nonumber \\[2mm]
 && +\frac{\stackrel{\sim}{g}_{0}\ket{N+1}J^{2}\times\bra{\ell}g_{0}\ket{\ell}\bra{N+1}
 \stackrel{\sim}{g}_{0}}{1-J^{2}
 \times\bra{N+1}\overset{\sim}{g}_{0}\ket{N+1}\bra{\ell}g_{0}\ket{\ell}}\nonumber \\[2mm]
 && -\frac{J\stackrel{\sim}{g}_{0}\ket{N+1}\bra{\ell}g_{0}+Jg_{0}\ket{\ell}\bra{N+1}
 \stackrel{\sim}{g}_{0}}{1-J^{2}\times\bra{N+1}
 \overset{\sim}{g}_{0}\ket{N+1}\bra{\ell}g_{0}\ket{\ell}},
\end{eqnarray}
where
\begin{equation}
\overset{\sim}{g}_{0}(z)=\frac{2}{S+1}\sum_{n=1}^{S}\frac{\ket{n}\bra{n}}{z+2J\cos(\frac{n\pi}{S+1})}.\nonumber
\end{equation}
The same as the previous  discussion, one can neglect the perturbation term of $\Delta(z_0)$ if $J\gg 1$. Then except for the remaining  energy levels of $H_{\mathrm{M}}$, the other  energy levels of the system for the side chain  with $S$ points are obtained
\begin{equation}\label{eq:grmore}
\begin{cases}
\left<\ell|g_{0}(z_{0})|\ell\right>=0, & S=odd\\[2mm]
\left<\ell|g_{0}(z_{0})|\ell\right>+\displaystyle\frac{\lambda_{S}}{z_{0}}=0, & S=even
\end{cases}
\end{equation}
where, $\lambda_{S}=\displaystyle\frac{1}{2(S+1)}\sum_{n=1}^{S}\tan^{2}\displaystyle\frac{n\pi}{S+1}$.
We can find that if the number of points on the side chain is odd, the energy levels of the system are the same as the case of $S=1$ which has been  discussed in Subsec.~\ref{subsec:onepoint}. This implies that the distribution of the energy levels for the cases of side chain with odd points are the same. Then  propagation features of the particle in the triple graph for the side chain with odd points are analogous. So that all the system for the side chain with odd points can exhibit the switching effect when the number of points on the main chain and the position of the side chain satisfy concrete conditions.  Additionally, for the case of $S$ being even, the distribution of the energy levels resembles  the case of $S=2$ discussed in Subsec.~\ref{subsec:twopoint} due to the fact that the second equation in Eq.~(\ref{eq:grmore}) are of the same kind except for the difference of the coefficient $\lambda_S$. So the propagation features of the particle in the triple graphs for the side chain with even points resemble each other and always do not exhibit the switching effect, which is shown in Fig.~\ref{fig:twopoint} (a) and Fig.~\ref{fig:morepoint} (b). According to the above discussion, we can find that whether the particle can go through  the connection points into the other part of the main chain can be significantly  influenced by the parity of the number of points on the side chain, which differs  from the case studied in ref.~\cite{FAR2}. In that reference, the quantum walk on a chain of points with an attached NAND tree was studied, where the transmission coefficient  is determined by the evaluation results of the  NAND tree. Additionally the transmission coefficient can only be zero or one. In the system we considered, we show that the probability of the particle in one part of the main chain can oscillate with time for the side chain with even points.

\section{Experiment proposal and Conclusion}

Since the quantum-walk system we considered can exhibit the above novel features, we expect that can be observed in experiment and even be designed as switching device. Then we propose the following possible experimental proposals.
Firstly, we expect to realize such a quantum-walk system in solid systems. Assume there is a one-dimensional lattice system and  the particle with spin can tunnel between nearest sites. If one applies a external  magnetic field $B$ into the lattice system, the spin will be polarized to the direction of the magnetic field when the temperature is low enough. At this case, the spin degree of the particle is frozen, and then the Hamiltonian describing this system amounts to that of quantum walk in one-dimensional lattice without the side chain. Meanwhile, if one  applies another magnetic  field whose value and direction is equal and opposite to that of $B$, respectively, into the site $\ell$,   the total magnetic field at the site $\ell$ is zero. Then once the particle tunnel into the site $\ell$, the spin degree of the particle will be released so that there is probability for the spin of the particle  flipping into the another direction at the site $\ell$. Assuming the particle is an electron with spin being $1/2$, one can easily find that the Hamiltonian of this solid system is essentially equal to Eq.~(\ref{eq:model}) with $S=1$. So such a system is possible to realize the quantum walk in the triple graph with the side of one point.  Naturally, if the particle with spin larger than $1/2$,  such system can also be used to realize the quantum walk in the triple graph with the side of more points.  Note that the conventional solid system is difficult to simulate the quantum walk in the triple graph with side chain of more points because  the spin of the particle in conventional solid system  is usually equal to $1/2$ and  difficult to reach larger values. Whereas,  the cold-atom systems trapped in magnetic lattice  can make up for this defect due to the fact  that the spin of the particle can  more various for different alkali and alkali earth metals.

The aforementioned  quantum-walk system is also expected to be realized in artificial systems. For example, $N$-qubits are coupled by the single-mode resonant cavity, and the interaction Hamiltonian was given in ref.~\cite{Yangsy}. If at the initial time, one of the qubits is applied into a microwave pulse that  is resonant with the transition between two levels of the qubit, the initial state of the $N$-qubit system is $\prod_{k\neq j}\ket{\textrm{g}}_k\ket{\textrm{e}}_j\ket{0}_\textrm{c}$ . Such a initial state  means that the $j$-th qubit  is in its exited state $\ket{\textrm{e}}$, the others are in  ground state $\ket{\textrm{g}}$, respectively,  and the cavity mode is initially in the vacuum state $\ket{0}_\textrm{c}$. For convenience, the initial state of the system can be written as $\ket{\textrm{e}}_j\ket{0}_\textrm{c}$.
In second-order perturbation and taking the state $\prod_{k}\ket{\textrm{g}}_k\ket{1}_\textrm{c}$ as the medium state, the effective Hamiltonian just is the Eq.~(\ref{eq:model}) without the side chain, and the Fock basis corresponding to the system can be simply denoted as a  single-particle state $\ket{j}$ ($j=1,2, \cdots  N$). Additionally, one can set up  another $S$-qubits coupled by another single-mode resonant cavity  as the side chain, and meanwhile  couple the first qubit of the side chain with the $\ell$th qubit of the above $N$-qubits.  In this case, the system can simulate the quantum-walk in the triple graph we considered. Note that the coupling strength $J$ can be tuned  due to the device parameters and/or the placement of the qubits in the cavity can be changed. So the coupled-qubits system maybe more powerful to realize the quantum-walk in the triple graph we considered.

In the above, we investigated the quantum random walk on the triple type graph and show the effect of  the side chain on the  propagation features on the main chain.  We found that if the tunneling strength on the side chain is much larger than that on the main chain, the system can exhibit the switching effect  for the case of the side chain with odd points when the number of points on the main chain and the position of the side chain satisfy concrete conditions. Whereas, for the case of the side chain with even points, the particle can always pass through  the connection point into the other part of the main chain no matter where the connection point between the main chain and the side chain  is. So  the quantum walk on the triple type graph with the side chain of even points can not exhibit the switching effect, which is contrast to the case of the side chain with odd points.  Such propagation properties were  explained with the help of Green function. We also suggest two proposals for experiment to observe such an effect, which is expected to be enlightening for the design of new type of switching devices.

\section*{Acknowledgment}
The work is supported by the NBRP of China (2014CB921201), and the NSFC ( 11274272 and 11434008),  and by Fundamental Research Funds for Central Universities.


\begin{thebibliography}{22}
\expandafter\ifx\csname natexlab\endcsname\relax\def\natexlab#1{#1}\fi
\expandafter\ifx\csname bibnamefont\endcsname\relax
  \def\bibnamefont#1{#1}\fi
\expandafter\ifx\csname bibfnamefont\endcsname\relax
  \def\bibfnamefont#1{#1}\fi
\expandafter\ifx\csname citenamefont\endcsname\relax
  \def\citenamefont#1{#1}\fi
\expandafter\ifx\csname url\endcsname\relax
  \def\url#1{\texttt{#1}}\fi
\expandafter\ifx\csname urlprefix\endcsname\relax\def\urlprefix{URL }\fi
\providecommand{\bibinfo}[2]{#2}
\providecommand{\eprint}[2][]{\url{#2}}


\bibitem{Ahar}
Y. Aharonov, L. Davidovich, and N. Zagury, Phys. Rev. A, {\bf{48}},1687 (1993).

\bibitem{Moh}
 M. Mohseni, P. Rebentrost, S. Lloyd, and A. Aspuru-Guzik, J. Chem. Phys, {\bf{129}}, 174106 (2008).

 \bibitem{Shen}
 N. Shenvi, J. Kempe, and R.B. Whaley, Phys. Rev. A, {\bf{67}},052307
 (2003).

 \bibitem{Sal}
 S. E. Venegas-Andraca, Quantum Information Processing {\bf{11}}, 1015 (2012).


\bibitem{Far}
E. Farhi and S. Gutmann, Phys. Rev. A, {\bf{58}}, 915 (1998).


\bibitem{Kem}
J. Kempe, Probability Theory Relat.Fields, {\bf{133}}(2), 215 (2005).

\bibitem{Kos}
J. Ko\v{s}\'ik and V. Bu\v{z}ek, Phys. Rev. A, {\bf{71}}, 012306 (2005).

\bibitem{Kro}
H. Krovi and T. A. Brun, Phys. Rev. A, {\bf{73}}, 032341(2006).

\bibitem{Mar}
F. L. Marquezino, R. Portugal, G. Abal, and R. Donangelo, Phys. Rev. A, {\bf{77}}, 042312 (2008).


\bibitem{Adi}
A. Makmal, M. Zhu, D. Manzano, M. Tiersch, and H. J. Briegel, Phys. Rev. A, {\bf{90}}, 022314 (2014).


\bibitem{c1}
M. Bednarska, A. Grudka, P. Kurzy\'nski, T. \L{}uczak and A. W\'ojcik, Phys. Lett. A, {\bf{317}}, 21 (2003).

\bibitem{c2}
D. Solenov and L. Fedichkin, Phys. Rev. A, {\bf{73}}, 012313 (2006).

\bibitem{c3}
D. D'Alessandro, G. Parlangeli, and F. Albertini, Jour. Phys. A, {\bf{40}}, 14447 (2007).




\bibitem{hcy1}
D. Solenov and L. Fedichkin, Phys. Rev. A, {\bf{73}}, 012308 (2006).

 \bibitem{p1}
 G. Leung, P. Knott,  J. Bailey and V. Kendon, New. Jour. Phys. {\bf{12}}, 123018 (2010).



\bibitem{p2}
B. Koll\'ar, T. Kiss, J. Novotn\'y, and I. Jex, Phys. Rev. Lett, {\bf{108}}, 230505 (2012)



\bibitem{p3}
B. Koll\'ar, J. Novotn\'y, T. Kiss and I. Jex, New Jour. Phys, {\bf{16}}, 023002 (2014).





\bibitem{gi1}
B. L. Douglas, J. B. Wang, Jour. Phys. A, {\bf{41}}, 075303 (2008).


\bibitem{gi2}
 J. K. Gamble, M. Friesen, D. Zhou, R. Joynt, and S. N. Coppersmith, Phys. Rev. A, {\bf{81}}, 052313 (2010).

\bibitem{gi3}
S. D. Berry and J. B. Wang, Phys. Rev. A, {\bf{83}}, 042317 (2011).


\bibitem{gi4}
K. Rudinger, J. K. Gamble, M. Wellons, E. Bach, M. Friesen, R. Joynt, and S. N. Coppersmith, Phys. Rev. A, {\bf{86}}, 022334 (2012).

  \bibitem{Yangsy}
  C. P. Yang, and S. Y. Han, Phys. Rev. A, {\bf{72}}, 032311 (2005).

\bibitem{FAR2}
E. Farhi, J.Goldstone, S.Gutmann, - eprint arXiv:quant-ph/0702144 - 2007
\bibitem{Ho}
Ho-Fai Cheung, Yuval Gefen, Eberhard K.Riedel, and Wei-Heng Shih, Phys. Rev. B,{\bf{37}}, 6050 (1988)

\end{thebibliography}
\end{document}